\documentclass[aps,prd,showpacs,twocolumn,amssymb,floatfix,nofootinbib,superscriptaddress]{revtex4-1}
\usepackage{graphicx,amsmath,subfigure,psfrag,times}
\usepackage{multirow} 
\usepackage{xspace}
\usepackage{bm,url}
\usepackage{enumerate}

\newcommand{\eeq}{\end{equation}}
\newcommand{\bea}{\begin{eqnarray}}

\def\ltsima{$\; \buildrel < \over \sim \;$}
\def\simlt{\lower.5ex\hbox{\ltsima}}
\def\gtsima{$\; \buildrel > \over \sim \;$}
\def\simgt{\lower.5ex\hbox{\gtsima}}

\def\lesssim{\mathrel{\hbox{\rlap{\hbox{\lower4pt\hbox{$\sim$}}}\hbox{$<$}}}}
\def\gtrsim{\mathrel{\hbox{\rlap{\hbox{\lower4pt\hbox{$\sim$}}}\hbox{$>$}}}}
\def\alt{\mathrel{\hbox{\rlap{\hbox{\lower4pt\hbox{$\sim$}}}\hbox{$<$}}}}
\def\agt{\mathrel{\hbox{\rlap{\hbox{\lower4pt\hbox{$\sim$}}}\hbox{$>$}}}}
\def\PRD{{\it Phys. Rev.} D~}

\def\apjl{{\it Astrophys. J.} Lett~}

\def\aap{{\it A\&A~ }}

\def\mnras{{\it MNRAS}} 
\def\nat{{\it Nature}} 
\def\gta{\ifmmode {\mathbin{\lower 3pt\hbox   
    {$\,\rlap{\raise 5pt\hbox{$\char'076$}}\mathchar"7218\,$}}}
    \else {${\mathbin{\lower 3pt\hbox
    {$\rlap{\raise 5pt\hbox{$\char'076$}}\mathchar"7218\,$}}}
    $}\fi}
\def\lta{\ifmmode {\,\mathbin{\lower 3pt\hbox   
    {$\,\rlap{\raise 5pt\hbox{$\char'074$}}\mathchar"7218\,$}}}
    \else {${\mathbin{\lower 3pt\hbox
    {$\rlap{\raise 5pt\hbox{$\char'074$}}\mathchar"7218\,$}}}
    $}\fi}

\newcommand{\SU}{\affiliation{Department of Physics, Syracuse University, Syracuse, NY 13244, USA}}
\newcommand{\KITP}{\affiliation{Kavli Institute for Theoretical Physics, University of California, Santa Barbara, CA 93106}}
\newcommand{\LL}{\affiliation{LIGO Laboratory, California Institute of Technology, Pasadena, CA 91125}}
\newcommand{\TAP}{\affiliation{Theoretical Astrophysics 130-33, California Institute of Technology, Pasadena, CA 91125}}

\begin{document}
\title{Effect of eccentricity on binary neutron star searches in Advanced LIGO}

\author{E. A. Huerta}\SU \KITP%
\author{Duncan A. Brown}\SU \KITP\LL\TAP%


\date{\today}

\begin{abstract}        
Binary neutron stars (BNSs) are the primary source of gravitational waves for
the Laser Interferometer Gravitational-wave Observatory (LIGO) and its
international partners Virgo and KAGRA. Current BNS searches target field
binaries whose orbits will have circularized by radiation reaction before
their gravitational waves enter the Advanced LIGO sensitive band at $15$~Hz.
It has been suggested that a population of BNSs may form by $n$-body
interactions near supermassive black holes or in globular clusters and that
these systems may have non-negligible eccentricity in the Advanced LIGO band.
We show that for BNS systems with total mass of $2.4\,
M_\odot$ ($6\, M_\odot$), the effect of eccentricity $e \lesssim 0.02$ $(0.05)$
is negligible and a circular search is effectual for these binaries.  For
eccentricities up to $e = 0.4$, we investigate the selection bias caused by
neglecting eccentricity in BNS searches. If such high eccentricity systems
exist, searches that specifically target eccentric binaries will be needed in
Advanced LIGO and Virgo.

\end{abstract}


\maketitle

\section{Introduction}   

Binary neutron star (BNS) mergers are expected to be rich sources of gravitational waves (GWs)~\cite{Th300} and the progenitors of short gamma-ray bursts (SGRBs)~\cite{Kochanek:1993mw}, among other electromagnetic counterparts~\cite{Metzger:2012}. These sources may be detected by advanced GW observatories~\cite{aLIGo, virgo,kagra} within a horizon distance of  445 Mpc~\cite{ratespap}. Observations of galactic double NS systems containing pulsars, which formed through binary evolution in the disk of the galaxy, suggest that five of these systems are expected to merge through GW radiation within a Hubble time~\cite{bulic, mandelbns}.  Observed field BNSs have orbital periods of a few hours,  merger times \(\sim 10^{2-3}\) Myr, low-kick velocities \(\lesssim 50 \,{\rm km \, s^{-1}}\), and are expected to enter the frequency band of ground-based GW detectors with negligible residual eccentricities~\cite{bulic}: the binary pulsar PSR 1913+16 is expected to have an eccentricity \(e \sim 10^{-6}\) when its gravitational-wave frequency is 15Hz~\cite{peters,Kalo:2001,bulic}.  Population synthesis models constrained by BNSs in the field predict a coalescence rate in Advanced LIGO (aLIGO) of 0.4--400 per year, with a likely value of 40 per year~\cite{ratespap}.

In addition to the observed field binaries, observations by Swift have detected a population of SGRBs in or near elliptical galaxies~\cite{Bart:2005}. Some of these SGRBs have spatial offsets that cannot be attributed to the ejection of the progenitor compact binary from the host galaxy by the kick imparted to the system through the supernova explosions that create its components~\cite{grind}. The necessary kick velocities for the progenitor compact binary to remain bound to the elliptical host galaxy and at the same time have its apogalacticon in the halo are very different to those observed in field BNSs~\cite{Berger:2005}. In~\cite{grind}, Grindlay et al. propose that SGRBs observed in or near ellipticals may be produced by the mergers of BNSs formed by dynamical capture in core-collapsed globular clusters. In these dense stellar environments, \(n\)-body interactions may lead to the formation of binaries that have sizable eccentricities at merger~\cite{east:2012}. Based on the globular-cluster population of elliptical galaxies, and scaling from the BNS observed in the globular cluster M15~\cite{Jac:2006},  Grindlay et al. showed that around  \(10-30\%\) of SGRBs may be  generated from BNSs in globular clusters, although the expected detection event rate of BNSs in globulars for ground-based detectors is a factor \(\sim 200\) smaller than the expected rate for field BNSs~\cite{ratespap,Piran:2006}.

Dynamical interactions in galactic nuclei may also lead to the formation of eccentric BNSs~\cite{Leary:2009, antonini}. BNSs formed near galactic nuclei may enter the LIGO band before the maximum eccentricity expected from Kozai oscillations is reached. After multiple periodic oscillations in the orbital elements, and before gravitational radiation becomes important, these BNSs may enter the LIGO band with high eccentricities~\cite{antonini}. Nonetheless, the contribution of these type of sources to the BNS coalesce rates is likely to be small, namely: 10, 1 and 0.1\(\%\) of the low, realistic and high estimates in~\cite{ratespap}.  The studies described in~\cite{Bart:2005,Leary:2009, antonini} do not take into account natal kicks. In~\cite{Chaurasia:2005}, it is shown that if a BNS system receives a supernova kick which is similar in magnitude to its circular speed, then the likelihood of undergoing an extremely fast coalescence is very high. Indeed, for BNS systems that experience planar kicks, about \(24\%\) of them coalesce ten thousand times faster than a BNS system that stays in a circular orbit immediately after the supernovae. These type of events would be observed as a supernova followed by an extremely fast coalescence that takes the form of a second consecutive catastrophic event. Since both events may take place in the same location within months or years apart, these events may be effectively detected as double supernovae. Pulsar surveys may not detect these events because current search algorithms are not equipped to cope with the strong Doppler smearing effects present in these systems~\cite{Chaurasia:2005}. A pulsar can be detected if its lifespan is of the order of 10 Myrs, and so highly eccentric systems which are driven to coalescence within 1Myr or less constitute a currently  `missing' BNS population that  aLIGO may detect ~\cite{Chaurasia:2005}.

Motivated by the possibility of an additional population of eccentric BNSs as GW sources, in this Brief Report, we revise the
study carried out in~\cite{Brown:2010} for aLIGO. Although
Ref.~\cite{Brown:2010} briefly considered eccentric binaries in the context of aLIGO, the computational cost of such an analysis for aLIGO's improved low-frequency sensitivity limited their study to only 280 simulated
signals in the parameter space they considered for Initial LIGO ($2 \le M /
M_\odot \le 13$ and $0 \le e \le 0.4$). The resolution of Figs. 8 and 9 in
Ref.~\cite{Brown:2010} is insufficient to understand the effect of
eccentricity in aLIGO searches. Hence, we have improved the numerical
techniques used to simulate the eccentric waveforms using the model introduced in~\cite{hinder}, and are now able to simulate
40,000 signals in the space of eccentric BNS signals. This allows us to
determine the eccentricity at which circular templates will fail and to
demonstrate that effort will be needed to construct new searches for low to
moderate eccentricities for advanced GW searches.  This work is complementary to that
of Ref.~\cite{Bence:2012}, which considers ``repeated bursts''
from highly eccentric binaries. 
Based on results of radio surveys that have  confirmed the existence of pulsars with masses heavier than the canonical value of  \(1.35 M_{\odot}\)~\cite{Thorsett:1999}, we consider binaries with component masses between  \(1 M_{\odot}-3 M_{\odot}\). Throughout this analysis, we use the Zero Detuned High Power (ZDHP) noise curve for aLIGO~\cite{ZDHP:2010}. The results of this study are organized as follows: In Section~\ref{bss} we describe the method used to construct a template bank of non-eccentric BNS waveforms and the model used to simulate eccentric BNS signals. Section~\ref{res} presents the results of our investigations. Finally, Section~\ref{conclu} presents a summary of our findings and future directions of work.


\section{Bank simulation studies}
\label{bss}

We now discuss the efficacy with which a template bank~\cite{Sathyaprakash:1991mt,Balasubramanian:1995bm} of non-eccentric waveforms can recover GW signals of eccentric BNSs. To do so, we introduce a few data-analysis concepts: on the vector space of signals, the Fourier transform of a GW signal \(h(t)\) is given by

\begin{equation}
\tilde{h} (f) =\int_{-\infty}^{\infty} e^{-2i\pi f t } h(t)\mathrm{d}t \, .
\label{ft}
\end{equation} 

The noise-weighted inner product between two signals \(h_1\) and \(h_2\) is given by

\begin{equation}
\left(h_1|h_2\right)= 2\int_{f_{\rm{min}}}^{f_{\rm{max}}} \frac{\tilde{h}_1^{*} (f) \tilde{h}_2 (f) + \tilde{h}_1(f) \tilde{h}_2^{*}(f)}{S_n(f)} \mathrm{d}f\, ,
\label{ip}
 \end{equation}  
 
\noindent where \(S_n(f)\) is the power spectral density (PSD) of the detector noise, i.e., aLIGO ZDHP. We set the lower limit of the integral in Eq. (2) to \(f_{\rm{min}}=15\)Hz. Given that the waveforms used in this study do not capture the merger and ring-down of the signals, we terminate the waveforms at the Schwarzschild innermost stable circular orbit (ISCO), namely, \(f_{\rm{max}} = f_{\rm{ISCO}} =  1/(6\sqrt{6} \pi M)\), where \(M\) stands for the total mass of the binary system. From this, we can construct the matched-filter signal-to-noise ratio (SNR)~\cite{Wainstein}

\begin{equation}
\rho = \frac{(s|h)}{\sqrt{(h|h)}}\, ,
\end{equation}

\noindent where $s = n (+ h)$ is the output of the detector, and $n$ is the detector noise. The normalized overlap between any two waveforms  \(h_1\), \(h_2\) is 

\begin{equation}
\left(\hat{h}_1 | \hat{h}_2\right)=   \frac{\left(h_1|h_2\right)}{\sqrt{ \left(h_1|h_1\right)\left(h_2|h_2\right) }}\, .
\label{over}
 \end{equation} 
 
 The time  of coalesce \(t_c\) and phase of coalescence \(\phi_c\)  are maximized over in both the SNR and the overlap~\cite{Allen:2005fk}. The maximized overlap
 
\begin{equation} 
 {\cal{O}} \left(h_1,  h_2\right) =  \underset{t_c,\, \phi_c }{\rm max}\, \left(\hat{h}_1|  \hat{h}_2  e^{i(2\pi f t_c - \phi_c)}  \right)\, ,
\label{mover}
 \end{equation}
 
\noindent gives the loss in signal-to-noise ratio incurred due to mismatches in the waveforms~\cite{FittingFactorApostolatos}. 
 To compute the effectualness of non-eccentric templates to recover eccentric systems, we construct a `bank' of templates which covers the space of circular BNS signals, and which are placed on a hexagonal lattice using the method introduced in~\cite{saton,BabaketalBankPlacement}, so that the loss in signal-to-noise ratio between a signal and the nearest template is no more than \(3\%\). The metric used to place the template grid is constructed using TaylorF2  inspiral templates that include corrections to the orbital phase evolution at 2 post-Newtonian (PN) order.  The waveform family used to create the template bank waveforms is Taylor T4~\cite{TaylorT4Origin,NRPNComparisonBoyleetal}, which includes corrections to the phase of the waveform to 3.5PN order~\cite{Blanchet:1995ez,Blanchet:2004ek,Blanchet:2005a}.  We have confirmed that this bank  is effectual for TaylorT4 waveforms before considering the effect of eccentricity. 

To simulate the eccentric BNS signals, we use the \(x\)-model introduced in~\cite{hinder}, which consists of a PN model that was calibrated by comparison to a numerical relativity (NR) simulation of an eccentric, equal-mass binary black hole. This model combines the 3PN conservative quasi-Keplerian orbit equations~\cite{Gopa:2004} with the 2PN evolution of the orbital elements~\cite{Gopa:2004b} to construct  an adiabatic waveform that best matches the NR simulation.  This model is named after  the choice of coordinates used to express the PN equations of motion, namely, the angular velocity of the compact objects, \(\omega\), via the variable \(x=\left(M\omega\right)^{2/3}\). This coordinate choice leads to a GW phase agreement between the PN model and the NR simulation used to calibrate the model of the order of \(\pm 0.1\) radians between 21 and 11 cycles before merger. Given the differences in the PN order for the energy flux between the \(x\)-model (2PN order) and TaylorT4 3.5 PN, we have studied the faithfulness between these two waveform families and have found that the faithfulness for a \((1.35 M_{\odot}, 1.35 M_{\odot})\)  BNS system is \(95\%\) in the limit \(e\rightarrow 0\). In order to carry out a detailed study of the importance of eccentricity in advanced searches of BNSs, we have substantially improved the implementation of the \(x\)-model described in~\cite{Brown:2010}, by adding an adaptive mesh-refinement method coupled with an accelerated integrator. 

As discussed in Ref.~\cite{Bence:2012}, PN calculations tend to slowly converge at late inspiral. Hence, more work is needed to improve the accuracy of PN-based waveforms for source detection and parameter estimation. This work is particularly important to accurately model systems with small pericenter distances and to faithfully capture the late inspiral evolution of eccentric systems. The \(x\)-model used to simulate the eccentric signals presents a similar behaviour: the phase difference between the \(x\)-model and the simulation used to calibrate it grows up to ~0.7 radians around 5 cycles before merger. This is a modeling issue that requires further improvement~\cite{hinder}. 

We compute the correlation or overlap between a simulated eccentric GW signal  (\(h^{e}\)) and the best fitting non-eccentric template (\(h^{T}_b\))  by maximizing the overlap over the templates in the bank. This is known as the Fitting Factor (\({\cal{FF}}\))~\cite{FittingFactorApostolatos}, and is defined as

\begin{equation}
{\cal{FF}} =   \underset{b\in {\rm{bank}} }{\rm max}\,  {\cal{O}}(h^{e}, h^{T}_b)\, .
\label{FF}
\end{equation} 

\section{Results}
\label{res}

The Monte Carlo simulation includes 40,000 points uniformly distributed in individual component mass \(1M_{\odot}<m_{1,\,2}<3M_{\odot}\), and with a uniform distribution of eccentricity in the range \(e\in[0,0.4]\).  The template waveforms are generated from an initial gravitational wave frequency of 15Hz.  We evolve the equations of motion (Eqs. (4)-(10) in~\cite{hinder}) of the eccentric waveform from a Keplerian mean orbital frequency of 7Hz (this Keplerian mean orbital frequency is one half the dominant, quadrupole GW frequency. Hence, the equations of motion of the \(x\)-model are evolved from a fiducial GW
frequency of 14Hz (see Eq. 3.2 in~\cite{Yunes:2009}) ).

\begin{figure*}[ht]
\centerline{
\includegraphics[height=0.38\textwidth,  clip]{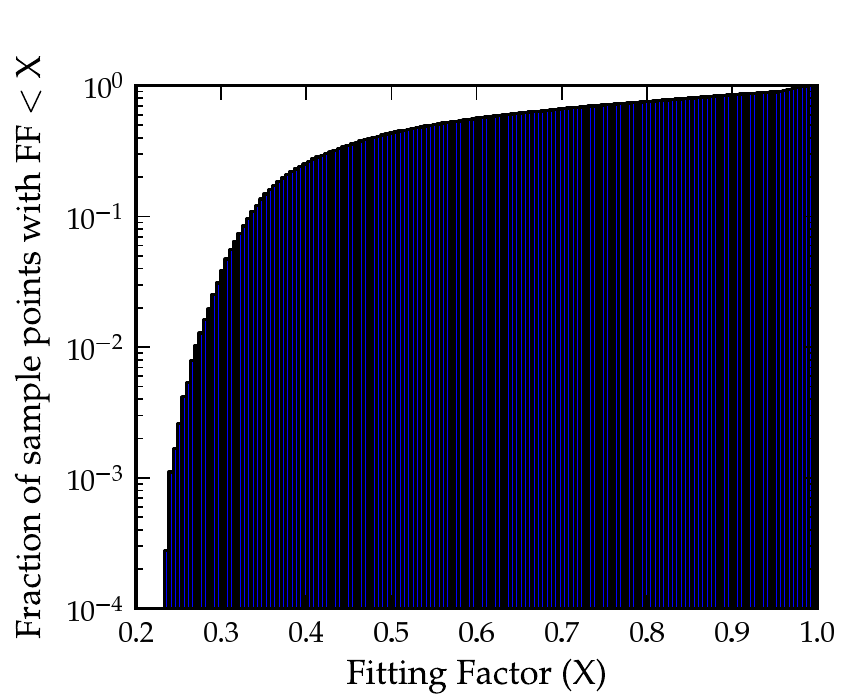}
\includegraphics[height=0.38\textwidth,  clip]{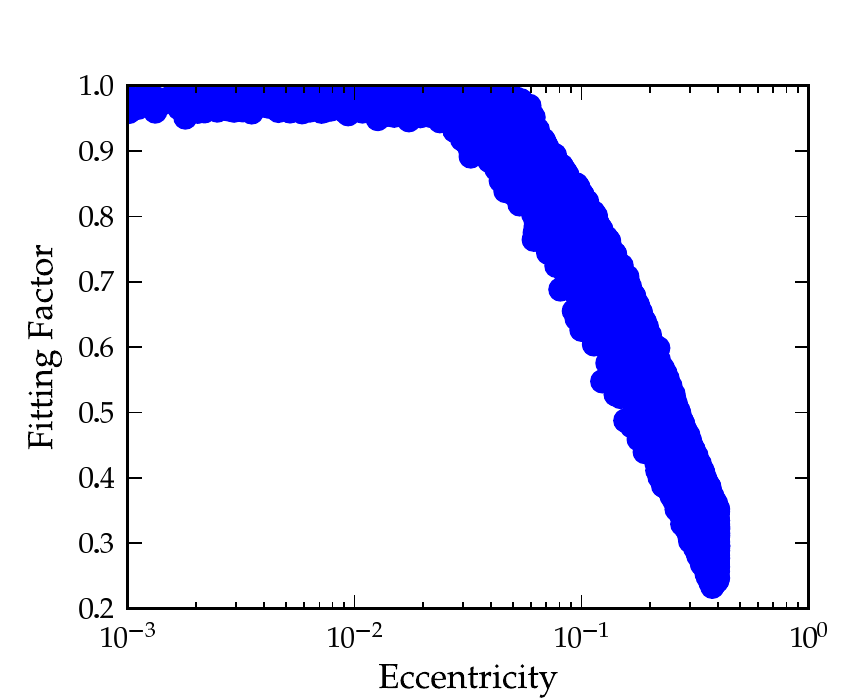}
}
\caption{The left panel is a cumulative histogram which indicates the fraction of points in the mass-space region where the bank of non-eccentric waveforms has a value of \({\cal{FF}}\)  less than the associated value on the x-axis. The right panel shows the \({\cal{FF}}\) as a function of the binary system's initial value of eccentricity.}
\label{FFI}
\end{figure*}

The left panel of Figure~\ref{FFI} presents a cumulative histogram of the injected eccentric signals with \({\cal{FF}}\) less than the value associated on the \(x\)-axis. There are two intrinsic parameters that determine the effectualness of the circular bank to recover eccentric signals. The right panel of Figure~\ref{FFI} shows that  the \({\cal{FF}}\)  decreases for increasing values of eccentricity, and suggests that some systems can be recovered with  \({\cal{FF}} \gtrsim 0.95\)  in an eccentricity range \(e\in[0.02,0.05]\).

\begin{figure*}[ht]
\centerline{
\includegraphics[height=0.38\textwidth,  clip]{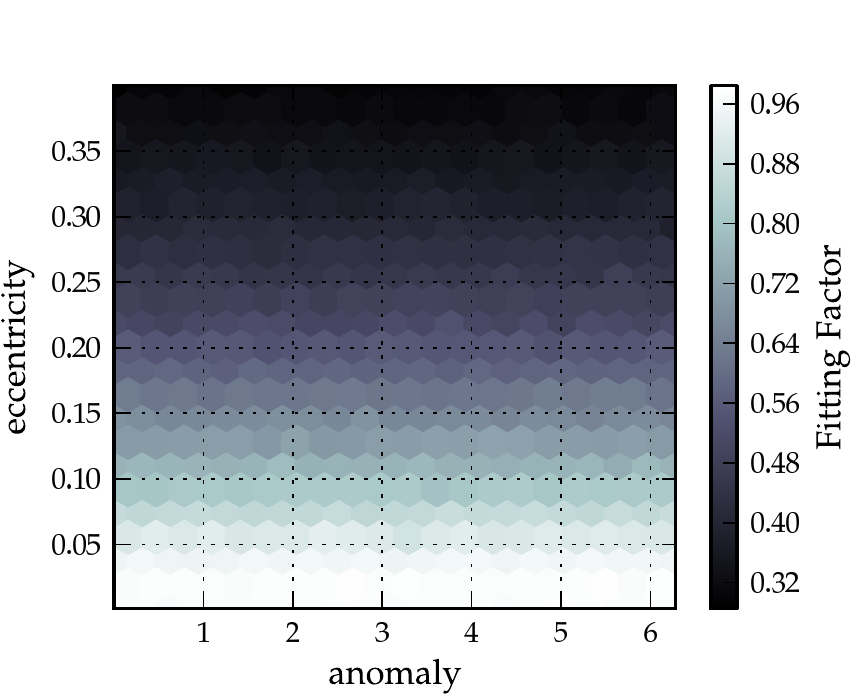}
\includegraphics[height=0.38\textwidth,  clip]{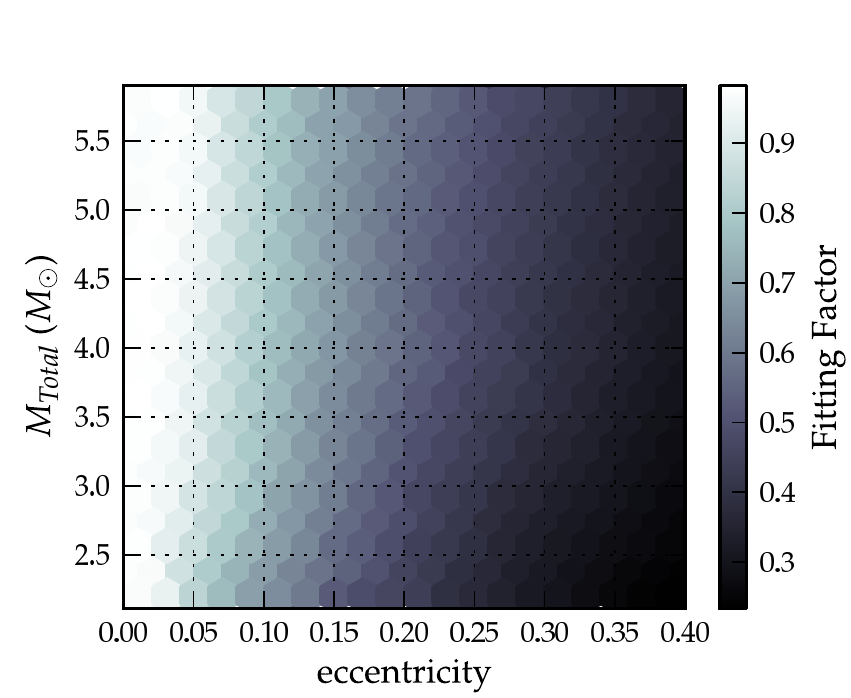}
}
\caption{The left panel shows the \({\cal{FF}}\) as a function of the initial value of the eccentricity and the anomaly of the trajectory for binaries with component masses  \(1M_{\odot} < m < 3M_{\odot}\).  The right panel shows the correlation of the \({\cal{FF}}\) with the initial orbital eccentricity and the total mass of the binary.}
\label{eanomFF}
\end{figure*}

Figure~\ref{eanomFF} shows that  the \({\cal{FF}}\)  depends primarily on the initial value of the eccentricity and on the total mass of the system. It does not depend on additional parameters that define the dynamics of the orbit at early inspiral, i.e., the value of the anomaly (left panel of Figure~\ref{eanomFF}), and depends only weakly on the mass ratio of the binary.  The right panel of Figure~\ref{eanomFF} shows that the  \({\cal{FF}}\) tends to be larger for systems with larger total mass, since  differences in GW phase evolution between eccentric and circularized waveforms have less cycles to accumulate.  In contrast, systems with low total mass live longer and the discrepancies in the waveforms accumulate over time, leading to larger phase discrepancies, and hence lower \({\cal{FF}}\)s. Therefore, using a template bank of circularized waveforms,  and the placement bank algorithm described in Section~\ref{bss}, is sufficient to recover signals with \(M_{\rm Total} \sim 2.4M_{\odot} \,(6M_{\odot})\) emitted by BNS systems whose eccentricity at a fiducial GW frequency of 14Hz is \(e\lesssim 0.02\, (0.05)\) with \({\cal{FF}}\)\(\gtrsim 0.95\).

We compare our findings with those reported in~\cite{Brown:2010}. 
Therein, it was found that Initial LIGO may be able to efficiently detect BNS systems with \(M_{\rm Total} \sim 2.4M_{\odot} \,(6M_{\odot})\) with residual eccentricity \(e\lesssim 0.05\, (0.08)\) at 40Hz. The authors in~\cite{Brown:2010} also probed the effect of eccentricity in the context of advanced LIGO assuming the PSD given in Eq. (29) of~\cite{Brown:2010} and using a low-frequency cut off of 10Hz. Given the demanding computational cost of such an analysis, they injected a few hundred signals to obtain a general, yet insufficient, picture of the efficacy with which aLIGO could recover eccentric BNS signals. Thus, in order to provide useful input to the ongoing planning of GW searches that will be carried out in the advanced detector era, we have substantially improved the analysis presented in~\cite{Brown:2010} for aLIGO. The results presented in this article, which are based on our up-to-date understanding of the sensitivity that advanced detectors will achieve, provide a complete picture of the efficacy with which circular templates can recovered signals from BNS systems with low to moderate eccentricity.

\section{Conclusions}
\label{conclu}

BNSs formed in core-collapsed globular clusters~\cite{grind} or near supermassive black holes in galactic nuclei~\cite{antonini}  may have high residual eccentricities when emitting gravitational radiation in aLIGO band. Still, the detection of eccentric BNSs  with the upcoming generation of GW detectors is not yet certain due to the uncertainty on the rate of these events~\cite{antonini}.  Nonetheless, the detection of eccentric BNSs is not implausible, and since the electromagnetic emission of these events is distinguishable from quasi-circular mergers, it is crucial to understand both types of events~\cite{east:2012}. The study presented in this paper is important to study the selection bias introduced  by neglecting eccentricity in advanced searches of BNSs, and  provides useful information when planning GW searches in the advanced detector era in order to extract the most from the observations.

The Monte Carlo analysis we have carried out shows that, assuming aLIGO ZDHP PSD,  BNS systems with total mass \( \sim 2.4M_{\odot} \,( 6M_{\odot})\)  and initial eccentricity \(e \lesssim 0.02\,(0.05)\) at a fiducial GW frequency of 14Hz, could be recovered with  \({\cal{FF}}\)\(\gtrsim 0.95\) using a template bank of circularized waveforms. Our findings suggests that in order to detect and study the rate of eccentric stellar-mass compact binaries in aLIGO, a search specifically targeting these systems will need to be constructed.

Extending this study to higher-mass systems systems requires input from NR simulations both to construct template banks and to simulate the signals. Recent simulations of eccentric neutron star-black hole (NSBH) systems show that GW emission is significantly larger than perturbative models suggests for periapsis distances close to effective innermost stable separations~\cite{east:2012a}. Modeling these events requires higher resolution simulations, additional computational resources, and a better method for creating initial data for eccentric binaries~\cite{east:2012a, Foucart:2012}. Furthermore, PN approximants are not suitable for detection purposes if the total mass of the system \(M\gtrsim 11.4 M_{\odot}\)~\cite{pnbuo, Prayush:2013a}.  Hence, we have considered binary systems that are not affected by waveform uncertainties in aLIGO band. The development of accurate waveform models to extend this analysis to higher-mass systems~\cite{ET1,ET2, Huerta:2012}, and the construction of a template matched filter search for eccentric binaries is the subject of  future work. 

\section*{Acknowledgments}
This work was supported by NSF grants PHY-0847611 and PHY-0854812. DB and EH would also like to thank the Research Corporation for Science Advancement and the Cottrell Scholars program for support. The Monte Carlo simulations described in this paper were performed using the  Syracuse University Gravitation and Relativity  (SUGAR) cluster,  which is supported by NSF grants PHY-1040231, PHY-0600953 and PHY-1104371. EH thanks Stefan Ballmer, Alessandra Buonanno, Peter Couvares, Jolien Creighton, Ryan Fisher, Chris Fryer, Ian Harry, Alex Huerta, Prayush Kumar, Alex Nitz, Bangalore Sathyaprakash, Peter Saulson and Matt West for useful discussions. We thank Ray Frey for carefully reading this manuscript. We also thank the anonymous referee for their constructive comments. Part of this work was carried out at the Kavli Institute for Theoretical Physics at Santa Barbara University, which is supported in part by the National Science Foundation under Grant No. NSF PHY05-51164

\bibliography{references}

\end{document}